\documentclass[paper]{JHEP}
\usepackage{epsfig}
\usepackage{amssymb}
\def\beq{\begin{equation}}
\def\beqn{\begin{eqnarray}}
\def\eeq{\end{equation}}
\def\eeqn{\end{eqnarray}}
\def\abs#1{\left|#1\right|}

\def\HW{{\small HERWIG}}
\def\DiracSlc#1{\slash\!\!\!#1}
\def\DiracSld#1{\slash\!\!\!\!#1}
\newcommand\sss{\scriptscriptstyle\rm}
\newcommand\bl{\overline{l}}

\newcommand\bu{\overline{u}}
\newcommand\bv{\overline{v}}
\newcommand\mV{m_{V}}
\newcommand\GammaV{\Gamma_{V}}
\newcommand\mVi{m_{V_i}}
\newcommand\GammaVi{\Gamma_{V_i}}
\newcommand\mW{m_{W}}
\newcommand\GammaW{\Gamma_{W}}
\newcommand\mt{m_t}
\newcommand\Gammat{\Gamma_t}
\newcommand\vecVcp{V_{Vl}}
\newcommand\axlVcp{A_{Vl}}
\newcommand\vecVicp{V_{V_il_i}}
\newcommand\axlVicp{A_{V_il_i}}
\newcommand\vecWcp{V_{Wl}}
\newcommand\axlWcp{A_{Wl}}
\newcommand\vecZcp{V_{Zl}}
\newcommand\axlZcp{A_{Zl}}

\newcommand\gw{g_{\sss W}}
\newcommand\thw{\theta_{\sss W}}
\newcommand\veps{\varepsilon}
\newcommand\Vtb{V_{tb}}
\newcommand\mydot{\!\cdot\!}
\newcommand\tM{\tilde{M}}

\newcommand\clH{{\mathbb H}}
\newcommand\clS{{\mathbb S}}
\newcommand\EVprjmap{{\cal P}_{\clH\to\clS}}

\newcommand\Gfun{{\cal G}}

\newcommand\Ms{M}
\newcommand\Ns{N}
\newcommand\As{A}
\newcommand\pt{p_{\sss T}}
\newcommand\kt{k_{\sss T}}
\newcommand\rp{r^\prime}

\preprint{
 Cavendish--HEP--07/01\hfill\\
 GEF--TH--09/2007\\
 ITP--UU--07/10\\
 NIKHEF/2007--004}
\title{Angular correlations of lepton pairs from vector boson
and top quark decays in\\ Monte Carlo simulations}
\author{Stefano Frixione\\
  INFN, Sezione di Genova,
  Via Dodecaneso 33, 16146 Genova, Italy\\
  E-mail: \email{Stefano.Frixione@cern.ch}}
\author{Eric Laenen\\
  NIKHEF,  Kruislaan 409, 1098 SJ Amsterdam, The Netherlands and\\
  Institute for Theoretical Physics, Utrecht University, 
  Leuvenlaan 4, 3584 CE Utrecht, The Netherlands\\
  E-mail: \email{Eric.Laenen@nikhef.nl}}
\author{Patrick Motylinski\\
  NIKHEF,  Kruislaan 409, 1098 SJ Amsterdam, The Netherlands\\
  E-mail: \email{patrickm@nikhef.nl}}
\author{Bryan R.\ Webber\\
  Cavendish Laboratory, %University of Cambridge\\
  J.J. Thompson Avenue, Cambridge CB3 0HE, U.K.\\
  E-mail: \email{webber@hep.phy.cam.ac.uk}}
\abstract{We explain how angular correlations in leptonic decays of vector
bosons and top quarks can be included in Monte Carlo parton showers, 
in particular those matched to NLO QCD computations. 
We consider the production of $n$ pairs of leptons, originating from the
decays of $n$ electroweak vector bosons or of $n$ top quarks, in the
narrow-width approximation. In the latter case, the information on the 
$n$ $b$ quarks emerging from the decays is also retained. We give results 
of implementing this procedure in MC@NLO.}

\begin{document}

\section{Introduction\label{sec:intro}}
Accurate predictions for the spectra of the leptons emerging from decays
of vector bosons or top quarks are important for a variety of studies
at hadron colliders, such as acceptance computations, tests of QCD,
and searches for new physics. Theoretical computations should be based 
on all Feynman diagrams in which the corresponding leptons are external
legs. In general, not all such diagrams are {\em resonant} diagrams, 
i.e. those in which the leptons directly emerge from a vector boson 
propagator (which, in the case of top decays, in turn is directly 
connected to the top quark via a $Wtb$ vertex). Usually, however,
predictions based on computations that retain only the resonant diagrams
are excellent approximations to those based on the fuller set of
diagrams, owing to the rather narrow widths of the vector bosons
and top quarks; the more so in the presence of final-state cuts which
are designed to enhance on-shell contributions.

A further approximation can be made, which we call the {\em decay chain}
approximation: resonant diagrams are replaced by diagrams relevant to 
the production of on-shell vector bosons or top quarks, times the diagrams
corresponding to the matrix elements for the decays. In this way, off-shell
effects are lost, but they can be recovered to some accuracy by reweighting
the results of the decay chain approximation by a Breit-Wigner function.
There is another piece of information that is lost in the decay chain
approximation, and cannot be recovered, namely that on {\em production
angular correlations} (more precisely, angular correlations due to production 
spin correlations). Let us denote by $P$ the decaying particle (a vector
boson or a top in our case), and by $d_1$, $\ldots$, $d_n$ its decay 
products, and consider the hard process
\beq
a+b\;\longrightarrow\; P(\longrightarrow d_1+\cdots+d_n)+X\,,
\label{proc}
\eeq
with $X$ a set of final-state particles which may also contain other
decaying vector bosons or top quarks. The process of eq.~(\ref{proc}) is
said to have decay angular correlations if the matrix elements of the
corresponding resonant Feynman diagrams have a non-trivial 
dependence\footnote{We denote here a particle and its four-momentum
by the same symbol.}
on $(d_i\mydot d_j)$. Clearly, decay correlations are always present if
the particle $P$ has spin different from zero. The process of eq.~(\ref{proc}) 
has production angular correlations if its matrix elements have a non-trivial
dependence on $(d_i\mydot a)$, $(d_i\mydot b)$, or $(d_i\mydot X)$.
It is therefore clear that the decay chain approximation can account
for the decay correlations, but not for the production correlations.

The decay chain approximation has obvious advantages, leading to
much simpler computations (especially at higher orders) owing to
the reduced multiplicity of the final state. Still, it
is not acceptable if the spectra of the decay products must be
predicted with some accuracy. The aim of this paper is to 
introduce an approach to the computations of lepton spectra
as given by resonant diagrams, which uses the decay chain approximation
but also correctly accounts for production angular correlations. 
The method is primarily intended to be applied to parton shower Monte
Carlos, including those that implement NLO QCD corrections such as 
MC@NLO~\cite{Frixione:2002ik,Frixione:2003ei} or POWHEG~\cite{Nason:2004rx}.
The idea stems from the following observation: the matrix elements
computed with the resonant diagrams are bounded from above by the
matrix elements obtained by eliminating the decay products and
putting the parent particles (vector bosons and/or top quarks)
on-shell, times a process-independent constant. One can therefore 
use the latter matrix elements (which we call {\em undecayed} matrix
elements) to perform computing-intensive tasks for which production
correlations are not an issue. When the four-momenta of the parent
particles are available, the resonant diagrams (we refer to the
corresponding matrix elements as {\em leptonic} ones) are used in
the context of a simple hit-and-miss procedure to generate the
leptonic four-momenta.

In order to apply a hit-and-miss procedure, we need upper bounds on the decay
matrix elements that are universal with respect to the production process.
These are derived in the following section, first for vector boson, then for
top quark decay, and finally for final states containing several vector bosons
and/or top quarks. The practical application of these results is discussed in
section~\ref{sec:appl}. The inclusion of angular correlations in NLO
computations is hampered by the presence of virtual corrections and the 
necessity for subtraction terms, which mean that one has to deal with 
expressions that are not simply matrix elements squared, and therefore 
are not necessarily positive-definite. This implies that the scheme we propose 
in this paper is such that angular correlations are not accurate to NLO
in the whole phase space, but are correct to NLO for hard real emissions
and to LO in soft and collinear regions. Obviously, one can implement
angular correlations exactly to NLO accuracy by using lepton 
matrix elements in all the steps of the computation. In this
paper, however, we are solely interested in the decay chain approximation. 
Illustrative results of our approach, obtained with MC@NLO, are presented 
in section~\ref{sec:res}, followed by our conclusions in 
section~\ref{sec:concl}. An appendix presents an alternative derivation 
of the upper bound for vector boson decay, which may
clarify some of the assumptions involved.

\section{Upper bounds for the leptonic matrix elements\label{sec:upp}}
In this section, we derive the universal factors that, when multiplied
by the undecayed matrix elements, give an upper bound for the leptonic
matrix elements. After introducing some notation, we shall treat the
cases of the vector bosons and of the top quarks in turn.

\subsection{Notations}
We shall always denote by
\beq
V\;\;\longrightarrow\;\;l\bl
\eeq
the decay of the vector boson $V\equiv W$ or $Z$ into a
lepton-antilepton pair, which means that in the case of $W$ decay
$\bl$ is not the antiparticle of $l$. In our conventions, the $Vl\bl$ 
vertex is
\beq
-i F_V\gamma^\mu\left(\vecVcp-\axlVcp\gamma_5\right)\,,
\eeq
where
\beqn
&&F_Z=\frac{\gw}{2\cos\thw},\;\;\;\;
\vecZcp=V_l,\;\;\;\;
\axlZcp=A_l;\\
&&F_W=\frac{\gw}{2\sqrt{2}},\;\;\;\;
\vecWcp=1,\;\;\;\;
\axlWcp=1.
\eeqn
We shall consider the process
\beqn
a(P_1)+b(P_2)&\longrightarrow&
V_1(q_1)+\ldots +V_n(q_n)+X(x)
\label{vecproc}
\\
&\longrightarrow&
l_1(k_1)+\bl_1(k_2)+\ldots +l_n(k_{2n-1})+\bl_n(k_{2n})+X(x),
\label{lepproc}
\eeqn
where
\beq
q_i=k_{2i-1}+k_{2i}\,,
\label{decmom}
\eeq
and $X$ collectively denotes any particles not originating from a vector 
boson decay. It is particularly convenient to write the phase space of 
the final-state particles of eq.~(\ref{lepproc}) as follows
\beqn
&&d\Phi_{2n+1^{\!\star}}(P_1+P_2;k_1,\ldots,k_{2n},x)=
\nonumber 
\\*&&\phantom{d\Phi a}
d\Phi_{n+1^{\!\star}}(P_1+P_2;q_1,\ldots,q_n,x)\,
\prod_{i=1}^n d\Phi_2(q_i;k_{2i-1},k_{2i})\,
\frac{dq_i^2}{2\pi}\,.
\label{phspfact}
\eeqn
As the notation $1^{\!\star}$ suggests, we treat the particles $X$ as 
a single particle with mass-squared $x^2$ and four-momentum $x$, since the
individual four-momenta of the particles $X$ are irrelevant in what
follows. On the r.h.s. of eq.~(\ref{phspfact}), the two-body phase 
spaces account for the decays
\beq
V_i(q_i)\;\longrightarrow\; l_i(k_{2i-1})+\bl_i(k_{2i}).
\eeq
The factorization formula of eq.~(\ref{phspfact}) is exact: the
vector bosons are off-shell, and their virtualities $q_i^2$ 
(i.e., the invariant masses of the lepton pairs) are explicitly 
integrated over. This decomposition has an obvious physical
interpretation in the context of resonant diagrams.

In the case of processes involving top quarks, we shall deal with
\beqn
a(P_1)+b(P_2)&\longrightarrow&
t_1(p_1)+\ldots +t_n(p_n)+X(x)
\label{topproc}
\\
&\longrightarrow&
W_1(q_1)+b_1(r_1)+\ldots +W_n(q_n)+b_n(r_n)+X(x)
\\
&\longrightarrow&
l_1(k_1)+\nu_1(k_2)+b_1(r_1)+\ldots +l_n(k_{2n-1})+\nu_n(k_{2n})+b_n(r_n)
\nonumber \\&&
+X(x)\,,
\label{fullproc}
\eeqn
where $t$ can be either a top or an antitop. As in eq.~(\ref{phspfact}),
we can also write the exact phase-space factorization
\beqn
&&d\Phi_{3n+1^{\!\star}}(P_1+P_2;k_1,\ldots,k_{2n},r_1,\ldots,r_n,x)=
\nonumber 
\\*&&\phantom{d\Phi a}
d\Phi_{n+1^{\!\star}}(P_1+P_2;p_1,\ldots,p_n,x)\,
\prod_{i=1}^n d\Phi_3(p_i;k_{2i-1},k_{2i},r_i)\,
\frac{dp_i^2}{2\pi}\,,
\label{tphspfact}
\eeqn
with the three-body phase spaces on the r.h.s. accounting for the decays
\beq
t_i(p_i)\;\longrightarrow\; W_i(q_i)+b_i(r_i)
\;\longrightarrow\; l_i(k_{2i-1})+\nu_i(k_{2i})+b_i(r_i).
\eeq

\subsection{Vector boson decay\label{sec:V}}
We start by considering the production of one $l\bl$ pair, and we
neglect the $Z/\gamma$ interference. The amplitude for the 
process in eq.~(\ref{lepproc}) with $n=1$ is
\beq
A=M^\mu\frac{i}{q^2-\mV^2+i\mV\GammaV}
\left(-g_{\mu\nu}+\frac{q_\mu q_\nu}{\mV^2}\right)
\bu(k_1)(-i F_V)\gamma^\nu \left(\vecVcp-\axlVcp\gamma_5\right)
v(k_2)\,,
\label{sVBamp}
\eeq
where $\mV$ and $\GammaV$ are the mass and the width of the vector boson
respectively, and $M^\mu$ is the amplitude for the process
\beq
a(P_1)+b(P_2)\,\longrightarrow\,
V(q)+X(x)\,,
\label{sVBproc}
\eeq
$\mu$ being the Lorentz index associated with $V$; the polarization
four-vector of $V$ is not included in $M^\mu$. From eq.~(\ref{sVBamp})
we get (neglecting lepton masses)
\beqn
\sum_{\rm spins}\abs{A}^2&=&
M^\mu M^{*\rho}
\frac{\left(-g_{\mu\nu}+q_\mu q_\nu/\mV^2\right)
\left(-g_{\rho\sigma}+q_\rho q_\sigma/\mV^2\right)}
{(q^2-\mV^2)^2+(\mV\GammaV)^2}
\nonumber \\*&\times&
F_V^2\,{\rm Tr}\!\left[\left(\vecVcp^2+\axlVcp^2-2\vecVcp\axlVcp\gamma_5\right)
\DiracSlc{k}_1 \gamma^\nu \DiracSlc{k}_2 \gamma^\sigma\right].
\label{sVBampsq}
\eeqn
We now consider the narrow width approximation $\GammaV\to 0$. We have
\beq
\frac{1}{(q^2-\mV^2)^2+(\mV\GammaV)^2}\;\longrightarrow\;
\frac{\pi}{\mV\GammaV}\,\delta\left(q^2-\mV^2\right)\,.
\label{BWnw}
\eeq
The $\delta$ function, which puts the vector boson on shell, allows us
to write
\beq
\left(-g^{\mu\nu}+\frac{q^\mu q^\nu}{\mV^2}\right)=
\sum_\lambda \veps_\lambda^\mu\veps_\lambda^{*\nu},
\label{polsum}
\eeq
where $\veps_\lambda$ are the polarization four-vectors of the vector
boson. Using eq.~(\ref{polsum}), eq.~(\ref{sVBampsq}) becomes
\beq
\sum_{spin}\abs{A}^2=\frac{\pi}{\mV\GammaV}\sum_{\lambda\lambda^\prime}
\tM_\lambda \rho_{\lambda\lambda^\prime}
\tM^*_{\lambda^\prime}\,
\delta\left(q^2-\mV^2\right),
\label{sVBampsq2}
\eeq
where we defined
\beq
\tM_\lambda=M_\mu \veps_\lambda^\mu\,,
\label{sVBtamp}
\eeq
which is the amplitude for the process of eq.~(\ref{sVBproc}) for a given
vector boson polarization $\lambda$. We also define
\beq
\rho_{\lambda\lambda^\prime}=
F_V^2\,{\rm Tr}\!\left[\left(\vecVcp^2+\axlVcp^2-2\vecVcp\axlVcp\gamma_5\right)
\DiracSlc{k}_1 \DiracSlc{\veps}_\lambda^* 
\DiracSlc{k}_2 \DiracSlc{\veps}_{\lambda^\prime} \right]
\label{rhodef}
\eeq
which is, apart from the normalization, the decay density 
matrix\footnote{The density matrix is usually defined as the transpose 
of that in eq.~(\ref{rhodef}). See e.g. ref.~\cite{Merzbacher}.} 
of the vector boson. This quantity can be explicitly computed; here, 
we only present it in the form of a diagonal matrix
\beq
\rho_{\lambda\lambda^\prime}=
\left(U\rho^D U^*\right)_{\lambda\lambda^\prime}\,,
\eeq
where
\beq
\rho^D=2\mV^2 F_V^2\,{\rm diag}\left(0,(\vecVcp-\axlVcp)^2,
(\vecVcp+\axlVcp)^2\right).
\label{rhodiag}
\eeq
The cross section for the production of a lepton pair in the narrow
width approximation is therefore
\beqn
d\sigma_{l\bl}&=&\frac{1}{2s}
\frac{\pi}{\mV\GammaV}\sum_{\lambda\lambda^\prime}
\left(\tM U\right)_\lambda \rho_{\lambda\lambda^\prime}^D
\left(\tM U\right)^*_{\lambda^\prime}\,\delta\left(q^2-\mV^2\right)
\nonumber \\*&&\times
\frac{dq^2}{2\pi}\,
d\Phi_{1+1^{\!\star}}(P_1+P_2;q,x)\,
d\Phi_2(q;k_1,k_2)\,.
\label{dsigll}
\eeqn
Owing to the hermiticity properties of the density matrices, and to
the explicit form of eq.~(\ref{rhodiag}), eq.~(\ref{dsigll}) is a
positive-definite quadratic form in the space of the spin indices
of the vector boson. Thus
\beq
d\sigma_{l\bl}\le
\frac{1}{2s}\frac{\pi}{\mV\GammaV}
\max_\lambda\left(\rho_{\lambda\lambda}^D\right)|\tM|^2
\frac{1}{2\pi}\,d\Phi_{1+1^{\!\star}}\,d\Phi_2\,,
\label{fupp}
\eeq
where
\beq
|\tM|^2=
\sum_\lambda \tM_\lambda \tM^*_\lambda =
\sum_\lambda\left(\tM U\right)_\lambda \left(\tM U\right)^*_\lambda =
M^\mu M^{*\nu}\left(-g_{\mu\nu}+\frac{q_\mu q_\nu}{\mV^2}\right)\,,
\eeq
with $q^2=\mV^2$. Eq.~(\ref{fupp}) cannot be used as an upper bound for the 
matrix element of the process~(\ref{lepproc}), since the measures on the
two sides are different. Using eq.~(\ref{BWnw}), we can however
easily reinstate the $q^2$ integration by inserting
\beq
1=\int\! dq^2\,\frac{\mV\GammaV}{\pi}\,\frac{1}{(q^2-\mV^2)^2+(\mV\GammaV)^2}
\label{nwinv}
\eeq
on the r.h.s. of eq.~(\ref{fupp}). Furthermore, from eq.~(\ref{rhodiag})
we obtain:
\beq
\max_\lambda\left(\rho_{\lambda\lambda}^D\right)=
2\mV^2 F_V^2\,(\vecVcp+\axlVcp)^2\,,
\label{rhomax}
\eeq
which holds since $\vecVcp\axlVcp >0$ regardless of the identity 
of the lepton $l$. Therefore
\beq
\frac{1}{2s}\sum_{spin}\abs{A}^2\equiv
\frac{d\sigma_{l\bl}}{d\Phi_{2+1^{\!\star}}}\le
\frac{2\mV^2\, F_V^2\,(\vecVcp+\axlVcp)^2}
{(q^2-\mV^2)^2+(\mV\GammaV)^2}\,
\frac{|\tM|^2}{2s}\,,
\label{Allupp}
\eeq
which strictly speaking holds only when $q^2=\mV^2$, since all results
in this section are formally derived in the limit $\GammaV\to 0$.
More details on this, and the reason for keeping a formal dependence 
on $q^2$ in eq.~(\ref{Allupp}), will be given in appendix~\ref{sec:app}.
Eq.~(\ref{Allupp}) is the main result of this section. It states 
that, in the narrow width approximation, the lepton-pair cross section
has an upper bound, which is a universal factor times the cross section
for the production of the parent vector boson
\beq
\frac{d\sigma_V}{d\Phi_{1+1^{\!\star}}}=\frac{1}{2s}\,|\tM|^2\,.
\eeq

\subsection{Top decay\label{sec:t}}
Here, we consider the decay of a top quark
\beq
t(p)\;\longrightarrow\; W^+(q)+b(r)
\;\longrightarrow\; l^+(k_1)+\nu(k_2)+b(r);
\eeq
the treatment of the decay of an antitop is fully analogous. 
Other top quarks may be present in the final state, but their decays 
are of no interest for the moment, and will be ignored. 
The amplitude for the process in eq.~(\ref{fullproc}) is
\beqn
A&=&\bu(r)\Vtb\frac{\gw}{2\sqrt{2}}\gamma^\mu(1-\gamma_5)
\frac{\DiracSlc{p}+\mt}{p^2-\mt^2+i\mt\Gammat}M
\nonumber \\*&\times&
\frac{-g_{\mu\nu}+q_\mu q_\nu/\mW^2}{q^2-\mW^2+i\mW\GammaW}
\bu(k_2)\frac{\gw}{2\sqrt{2}}\gamma^\nu(1-\gamma_5)v(k_1)\,,
\label{stopamp}
\eeqn
where $M$ is the amplitude for the process
\beq
a(P_1)+b(P_2)\,\longrightarrow\,
t(p)+X(x)\,,
\label{stopproc}
\eeq
except for a spinor $\bu(p)$, which is not included. Therefore, 
$M=\Gamma u(K)$, with $\Gamma$ a combination of $\gamma$ matrices,
and $K$ the four-momentum of a fermion entering the hard scattering.
By squaring eq.~(\ref{stopamp}) we get
{\renewcommand\jot{8pt}
\beqn
\abs{A}^2&=&\frac{\gw^4\abs{\Vtb}^2}{64}
\frac{1}{(p^2-\mt^2)^2+(\mt\Gammat)^2}
\frac{1}{(q^2-\mW^2)^2+(\mW\GammaW)^2}
\nonumber\\*&\times&
\bu(r)\gamma^\mu(1-\gamma_5)\left(\DiracSlc{p}+\mt\right)MM^*\gamma^0
\left(\DiracSlc{p}+\mt\right)(1+\gamma_5)\gamma^\rho u(r)
\nonumber\\*&\times&
\bu(k_2)\gamma_\mu(1-\gamma_5)v(k_1)
\bv(k_1)(1+\gamma_5)\gamma_\rho u(k_2)\,.
\label{stopampsq}
\eeqn}$\!$
Following what was done in eq.~(\ref{sVBampsq}), we now consider 
eq.~(\ref{stopampsq}) in the narrow width approximation $\Gammat\to 0$, 
i.e. we make the replacement
\beq
\frac{1}{(p^2-\mt^2)^2+(\mt\Gammat)^2}\;\longrightarrow\;
\frac{\pi}{\mt\Gammat}\,\delta\left(p^2-\mt^2\right)\,.
\label{topnw}
\eeq
Thanks to the on-shell condition introduced in this way, we can use the 
analogue of eq.~(\ref{polsum})
\beq
\DiracSlc{p}+\mt=\sum_\lambda u_\lambda(p)\bu_\lambda(p),
\eeq
which in turn suggests introducing the quantity
\beq
\tM_\lambda=\bu_\lambda(p)M\;\;\;\Longrightarrow\;\;\;
\tM_\lambda^*=M^*\gamma^0 u_\lambda(p),
\eeq
which is the analogue of eq.~(\ref{sVBtamp}), and is the amplitude for 
the process of eq.~(\ref{stopproc}) for a given top polarization $\lambda$.
Upon summing over the spins of the final-state leptons and $b$ quark,
eq.~(\ref{stopampsq}) can be cast in the same form as eq.~(\ref{sVBampsq2}):
\beq
\sum_{spin}\abs{A}^2=\frac{\pi}{\mt\Gammat}\sum_{\lambda\lambda^\prime}
\tM_\lambda \rho_{\lambda\lambda^\prime}
\tM^*_{\lambda^\prime}\,
\delta\left(p^2-\mt^2\right),
\label{stopampsq2}
\eeq
with
{\renewcommand\jot{8pt}
\beqn
\rho_{\lambda\lambda^\prime}&=&
\frac{\gw^4\abs{\Vtb}^2}{16}
\frac{1}{(q^2-\mW^2)^2+(\mW\GammaW)^2}\,\,
{\rm Tr}\!\left[(1-\gamma_5)\DiracSlc{k}_2\gamma_\mu 
\DiracSlc{k}_1\gamma_\rho\right]
\nonumber \\*&\times&
\bu_{\lambda^\prime}(p)(1+\gamma_5)\gamma^\rho\DiracSlc{r}
\gamma^\mu u_\lambda(p)\,.
\label{trhodef}
\eeqn}$\!$
This is the decay density matrix for the top quark, the analogue of
eq.~(\ref{rhodef}). We can now proceed exactly as was done in 
sect.~\ref{sec:V}, and therefore we must compute the decay density 
matrix, diagonalize it, and find the largest of the matrix elements 
so obtained. An explicit computation leads to 
\beq
\rho^D=\frac{4\gw^4\abs{\Vtb}^2}
{(q^2-\mW^2)^2+(\mW\GammaW)^2}\,
(r\mydot k_2)(p\mydot k_1)\,{\rm diag}(0,1)\,.
\label{trhodiag}
\eeq
Using eq.~(\ref{trhodiag}) and reinstating the integral in $dp^2$
using the analogue of eq.~(\ref{nwinv}), we finally arrive at
\beq
\frac{1}{2s}\sum_{spin}\abs{A}^2\equiv
\frac{d\sigma_{l\nu b}}{d\Phi_{3+1^{\!\star}}}\le
\frac{4\gw^4\abs{\Vtb}^2 (r\mydot k_2)(p\mydot k_1)}
{\Big((q^2-\mW^2)^2+(\mW\GammaW)^2\Big)
\Big((p^2-\mt^2)^2+(\mt\Gammat)^2\Big)}\,
\frac{|\tM|^2}{2s}\,,
\label{Alnubupp}
\eeq
which strictly speaking holds only when $p^2=\mt^2$. Eq.~(\ref{Alnubupp})
is the analogue of eq.~(\ref{Allupp}), and expresses the upper bound
on the matrix elements for the production of $l\nu b$ in terms of the
matrix elements for the production of a top quark
\beq
\frac{d\sigma_t}{d\Phi_{1+1^{\!\star}}}=\frac{1}{2s}\,|\tM|^2\,.
\eeq
In contrast to eq.~(\ref{Allupp}),
the bound of eq.~(\ref{Alnubupp}) is not a constant over the phase space 
of the particles emerging from top decay, because of its dependence on
$(r\mydot k_2)$ and $(p\mydot k_1)$. This helps to increase the 
efficiency of event generation in the context of an unweighting procedure, 
but in order to avoid any biases the phase-space must be sampled in
such a way as to reproduce exactly the $q^2$-, $(r\mydot k_2)$-, and 
$(p\mydot k_1)$-dependences of the bound. An alternative approach is
that of finding a constant larger than or equal to the bound, which
can be done by finding the maximum of the combination of dot products
\beq
D=(r\mydot k_2)(p\mydot k_1)\,.
\label{Dcomb}
\eeq
Using the top rest frame to perform the relevant computations,
it is a matter of simple algebra to obtain
\beq
D\le D_{\max}(q^2)=\left\{
\begin{array}{ll}
\mt^4/16 & \phantom{aaaa}q^2\le\frac{\mt^2}{2}\,, \\
q^2(\mt^2-q^2)/4 & \phantom{aaaa}\frac{\mt^2}{2}<q^2\le\mt^2\,.
\end{array}
\right.
\label{Dmax}
\eeq
Note that $\mt^4/16\ge q^2(\mt^2-q^2)/4$ in the whole
$q^2$ range, and therefore one can always set $D_{\max}=\mt^4/16$; this
is seen to lead to a very marginal degradation of unweighting efficiency.
We have therefore
\beq
\frac{d\sigma_{l\nu b}}{d\Phi_{3+1^{\!\star}}}\le
\frac{4\gw^4\abs{\Vtb}^2 D_{\max}(q^2)}
{\Big((q^2-\mW^2)^2+(\mW\GammaW)^2\Big)
\Big((p^2-\mt^2)^2+(\mt\Gammat)^2\Big)}\,
\frac{d\sigma_t}{d\Phi_{1+1^{\!\star}}}\,.
\label{Alnubupp2}
\eeq

\subsection{Multiple decays\label{sec:nV}}
It is easy to generalize the formulae derived in the previous sections
to the cases in which one is interested in the decay products of several 
vector bosons and/or top quarks. Consider for example the process
of eq.~(\ref{lepproc}). An equation identical to eq.~(\ref{sVBamp}) 
holds, with the formal replacements
\beqn
M^\mu&\longrightarrow&M^{\mu_1\ldots\mu_n}\,,
\\
\frac{-g_{\mu\nu}+q_\mu q_\nu/\mV^2}{q^2-\mV^2+i\mV\GammaV}
&\longrightarrow&
\prod_{i=1}^n 
\frac{-g_{\mu\nu}+q_{i\mu} q_{i\nu}/\mVi^2}{q_i^2-\mVi^2+i\mVi\GammaVi}\,,
\\
F_V\bu(k_1)\gamma^\nu \left(\vecVcp-\axlVcp\gamma_5\right)v(k_2)
&\longrightarrow&
\prod_{i=1}^n 
F_{V_i}\bu(k_{2i-1})\gamma^\nu 
\left(\vecVicp-\axlVicp\gamma_5\right)v(k_{2i})\,.\phantom{aaa}
\eeqn
The analogue of eq.~(\ref{dsigll}) features the quantity
\beq
\sum_{\lambda_1\lambda_1^\prime}\ldots \sum_{\lambda_n\lambda_n^\prime}
\left(\tM U_1\ldots U_n\right)_{\lambda_1\ldots\lambda_n} 
\rho_{1\lambda_1\lambda_1^\prime}^D\ldots
\rho_{n\lambda_n\lambda_n^\prime}^D
\left(\tM U_1\ldots U_n\right)^*_{\lambda_1^\prime\ldots\lambda_n^\prime} \,,
\eeq
which results from the simultaneous diagonalization of the spin density
matrices of the $n$ vector bosons. This allows one to use eq.~(\ref{rhomax}),
and to proceed as in the previous section. We therefore arrive at
\beq
\frac{d\sigma_{l_1\bl_1\ldots l_n\bl_n}}{d\Phi_{2n+1^\star}}\le
\left(\prod_{i=1}^n \frac{2\mVi^2\, F_{V_i}^2\,(\vecVicp+\axlVicp)^2}
{(q_i^2-\mVi^2)^2+(\mVi\GammaVi)^2}\right)\,
\frac{d\sigma_{V_1\ldots V_n}}{d\Phi_{n+1^\star}}\,,
\label{fnupp}
\eeq
where $d\sigma_{V_1\ldots V_n}$ is the cross section for the process
of eq.~(\ref{vecproc}), all the vector bosons being on-shell. 

Along the same lines, eq.~(\ref{Alnubupp2}) can immediately be generalized 
to the case of the decays of several top and antitop quarks:
\beqn
&&\frac{d\sigma_{l_1\nu_1 b_1\ldots l_n\nu_n b_n}}{d\Phi_{3n+1^\star}}\le
\nonumber \\*&&\phantom{aaaa}
\left(\prod_{i=1}^n \frac{4\gw^4\abs{\Vtb}^2 D_{\max}(q_i^2)}
{\Big((q_i^2-\mW^2)^2+(\mW\GammaW)^2\Big)
\Big((p_i^2-\mt^2)^2+(\mt\Gammat)^2\Big)}\right)\,
\frac{d\sigma_{t_1\ldots t_n}}{d\Phi_{n+1^\star}}\,.\phantom{aaaa}
\label{ftopnupp}
\eeqn
Obviously, eqs.~(\ref{fnupp}) and~(\ref{ftopnupp}) can be combined for
the simultaneous presence of vector bosons and top quarks in the
final state.

\section{Angular correlations in MC@NLO\label{sec:appl}}
As mentioned in the introduction, the straightforward way to predict
correctly all features of lepton spectra is to include in the computation 
the leptonic matrix elements, for example as done in MC@NLO version
3.3~\cite{Frixione:2006gn} for the cases of single-$V$ or $V\!H$ production,
or in refs.~\cite{Campbell:2004ch,Cao:2004ky} for the case of top 
quark decay in parton-level pure NLO computations.
We remind the reader that, in the context of MC@NLO, parton-level cross
sections (which are obtained by suitably modifying those which enter 
pure-NLO computations) are first integrated over the phase space of 
the final-state particles. The information gathered in this integration step
is then used in the event-generation step, whose aim is that of obtaining a
set of kinematic configurations (the {\em hard events}), which are
subsequently showered by the parton shower Monte Carlo. We also point 
out that the same integration-and-generation structure is used by POWHEG 
(although the cross sections integrated in the two formalisms
are not the same). The integration time increases rapidly
with the number of final-state particles; there is a corresponding
decrease in the efficiency of the generation of hard events.
This is the reason why it is interesting to find alternative ways 
to predict angular correlations in large-multiplicity
processes. We stress that in principle, the implementation
in MC@NLO (or POWHEG) of a process with correct angular correlations is 
identical to that of the same process without such correlations.
The problem is a practical one, namely that production angular correlations
require the knowledge of the lepton matrix elements, and the increased
multiplicity with respect to the undecayed matrix elements entails loss 
of accuracy and generation efficiency.

The strategy we propose in this paper starts with the following steps.
\begin{itemize}
\item[1.] Integrate the undecayed matrix elements.
\item[2.] Generate hard events using the results of the previous step; thus,
vectors bosons and/or top quarks will be present in the final state,
but not their decay products.
\item[3.] For each hard event, generate (massless) lepton (and $b$ quark, 
in the case of top decays) four-momenta, uniformly within the decay phase 
space(s) of the corresponding parent particle(s).
\item[4.] Compute the lepton matrix element using the four-momenta
obtained in step 3, and the undecayed matrix element using the four-momenta
obtained in step 2. Generate a flat random number $r$. If the lepton
matrix element, divided by its upper bound as given in eqs.~(\ref{fnupp}) 
and~(\ref{ftopnupp}), is smaller than $r$, throw the lepton four-momenta
away, and return to step 3.
\item[5.] Otherwise, replace the vector bosons and top quarks 
by the set of their decay products. The resulting kinematic
configuration is the leptonic hard event that can be showered by
the Monte Carlo.
\end{itemize}
Steps 3 to 5 constitute a standard hit-and-miss procedure, which guarantees 
that the lepton spectra reconstructed with the four-momenta of the leptonic 
hard event (and subsequent shower) will be identical to those computed by 
a direct integration of the leptonic matrix elements.

It is clear that the integration step will be greatly simplified by this
procedure: the number of phase-space variables relevant to the undecayed
processes~(\ref{vecproc}) and~(\ref{topproc}) is $n_U=3(n+n_X)-4$, whereas
$n_V=3(2n+n_X)-4$ and $n_t=3(3n+n_X)-4$ for leptonic processes~(\ref{lepproc}) 
and~(\ref{fullproc}) respectively. On the other hand, one may doubt
that the efficiency for producing leptonic hard events is larger 
than in the case of a straightforward integration of the leptonic matrix
elements. In fact, the adaptive integration performed in step 1 will only
give information on the $n_U$ degrees of freedom of the undecayed processes.
However, using the phase-space decompositions of eqs.~(\ref{phspfact})
and~(\ref{tphspfact}), one associates the extra $n_V-n_U=3n$ and 
$n_t-n_U=6n$ degrees of freedom with the decay phase spaces. Since we
are considering here only resonant diagrams, the leptonic matrix
elements will be fairly smooth in these extra $3n$ and $6n$ degrees
of freedom, if the parametrizations of the decay phase spaces are
properly chosen (the obvious choice of using the rest frame of the
decaying particles is also an optimal choice from this point of view).
Therefore, all of the complications due to the presence of several
peaks in the matrix elements are dealt with in step 2. The unweighting
performed in step 4 does not require any sophisticated numerical
approach (i.e., a preliminary adaptive integration is not necessary)
in order to achieve a satisfactory efficiency.

For the procedure as outlined above to work, it is crucial that the
leptonic matrix elements can be bounded from above by the undecayed
matrix elements. In the derivations of sect.~\ref{sec:upp} we have
assumed that the density matrix is positive definite, which is the case,
{\em and} that the matrix elements involved can be expressed as the
modulus squared of an amplitude. This is certainly the case in the 
context of a tree-level computation, but it is not true for all
the contributions to an NLO cross section. In particular, the interference
between virtual and Born amplitudes is not positive-definite in
general. The modified subtraction procedure~\cite{Frixione:2002ik} 
introduced in the MC@NLO formalism also implies the presence of a
second quantity which is possibly not positive-definite, namely the
difference between the real matrix elements and the MC subtraction
terms. The presence of non-positive-definite contributions is what
prevents one from including angular correlations exactly to NLO accuracy
in the context of the decay chain approximation, as anticipated
in sect.~\ref{sec:intro}.

Before proceeding, we remind the reader that there are two classes of
MC@NLO hard events, defined according to their kinematics: $\clS$ ($\clH$)
events have the same number of initial- and final-state particles as
Born (real-emission) contributions. Thus, the number of final-state particles
of $\clH$ events is equal to that of $\clS$ events, plus one. For example,
in $W^+W^-$ production (eq.~(\ref{vecproc})) we have $(n,n_X)=(2,0)$ for
$\clS$ events, and $(n,n_X)=(2,1)$ for $\clH$ events. In $t\bar{t}$ and
single-top production (eq.~(\ref{topproc})), we have $(n,n_X)=(2,0)$
and $(n,n_X)=(1,1)$ for $\clS$ events, and $(n,n_X)=(2,1)$ and
$(n,n_X)=(1,2)$ for $\clH$ events respectively. POWHEG (and, for that
matter, any NLO computation) also outputs $\clS$ and $\clH$ events.

We now extend the procedure proposed in points 1 to 5 above to the case 
of NLO computations matched to parton shower simulations, as follows:
\begin{itemize}
\item Steps 1 and 2 are unchanged.
\item For each $\clS$ event, go through steps 3 to 5, using Born-level
results to compute lepton matrix elements and their upper bounds.
\item For each $\clH$ event, compute a quantity $\Gfun(\clH)$ as
explained below and generate a random number $\rp$. If $\rp\!\le\Gfun(\clH)$, 
go through steps 3 to 5, using real-emission results to compute lepton 
matrix elements and their upper bounds. If \mbox{$\rp\!>\Gfun(\clH)$}, 
define an $\clS$-type event with the projection $\EVprjmap(\clH)$, and
proceed as explained for $\clS$ events above.
\end{itemize}
The definition of a map $\EVprjmap$ is a necessary condition for the
matching between NLO results and parton shower simulations: for more
details see e.g. refs.~\cite{Frixione:2003ei,Nason:2006hf}.
This implies that such a map need not be defined specifically
for the purpose of including angular correlations into MC@NLO or POWHEG.
The quantity $\Gfun$ is a largely arbitrary smooth and continuous
function, that assumes values between 0 and 1, and tends to 0 (1)
in the soft/collinear (hard-emission) regions. The role of $\Gfun$
is simply to avoid computing real-emission matrix elements
in the phase-space regions where they diverge. In the context of
MC@NLO, functions with the same behaviour as $\Gfun$ need be introduced
in order to ensure local cancellation between real matrix elements
and MC counterterms (see e.g. app.~A.5 of ref.~\cite{Frixione:2002ik}
and app.~B of ref.~\cite{Frixione:2003ei}), and one obvious choice
is that of setting $\Gfun$ equal to one of these functions (or to
a combination of them). 

It should be clear that the proposal made here accounts for angular 
correlations to LO accuracy close to the soft and collinear regions, 
since there $\Gfun(\clH)\simeq 0$, and therefore $\clH$ events are
projected onto $\clS$ events, for which we only consider the Born
matrix elements in the hit-and-miss procedure\footnote{We 
remind the reader that the full NLO undecayed matrix elements 
are used in steps 1 and 2.}. On the other hand, in the 
hard emission region only real corrections contribute to the
cross section, and thus angular correlations are included exactly
to NLO accuracy\footnote{One should bear in mind that radiation from 
the decay products is not included here.}. Angular correlations resulting 
from an MC matched to an NLO computation and implementing the method proposed 
in this paper have therefore the same or a better accuracy than LO-based
Monte Carlos. We also stress that angular correlations are actually
fairly close to those computed exactly to NLO, for two reasons.
First, NLO corrections to spin correlations are generally small.
Second, although virtual corrections and subtracted terms are not
positive definite, their angular correlations arising from the 
contributions (if any) that are proportional to the Born matrix elements
can be included exactly in the computation following the method proposed 
here, since both sides of eqs.~(\ref{fnupp}) and~(\ref{ftopnupp}) then 
get multiplied by the same factor.

\section{Results\label{sec:res}}
The approach described in the previous section has been adopted to include 
production angular correlations in MC@NLO in the cases of $W^+W^-$ production
(since version 3.1) and of $t\bar{t}$ and single-$t$ production
(since version 3.3). In this section we present sample results for 
$t\bar{t}$ and single-$t$ production, at the LHC ($pp$ collisions
at $\sqrt{S}=14$ TeV) and at the Tevatron run II ($p\bar{p}$ collisions 
at $\sqrt{S}=1.96$ TeV). All the predictions given in this 
section have been obtained by using the MRST2002 default PDF 
set~\cite{Martin:2002aw}, and by setting $m_t=175$~GeV and
$\Gamma_t=1.7$~GeV. In the case of single-$t$ production, we
also reconstruct the accompanying jets, by means of the $\kt$-clustering 
algorithm~\cite{Catani:1993hr}, with $d_{cut}=100$~GeV$^2$. We include 
in the clustering procedure all final-state stable hadrons and photons.
For the sake of simplicity, we force $\pi^0$'s and all lowest-lying 
$b$-flavoured states to be stable in \HW. The jets are ordered in 
transverse momentum.

%%%%%%%%%%%%%%%%%%%%%%%%%%%%%%%%%%%%%%%%%%%%%%%%%%%%%%%%%%%%%%%%%%%
\begin{figure}[htb]
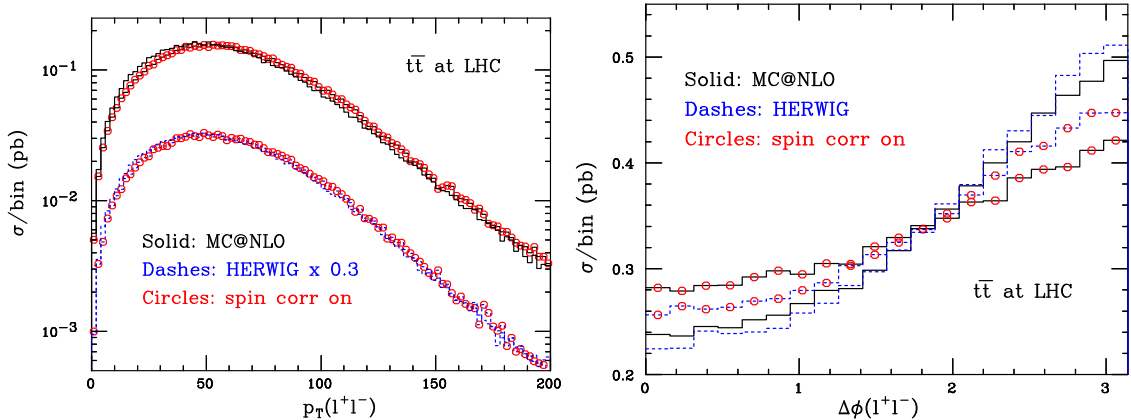

  \begin{center}
    \epsfig{figure=ptll_LHC2.eps,width=0.49\textwidth}
    \epsfig{figure=dphill_LHC2.eps,width=0.49\textwidth}
\caption{\label{fig:one} 
Transverse momentum of the lepton pair (left pane), and difference
in azimuthal angle between the leptons (right pane), in $t\bar{t}$ production 
at the LHC. \HW\ results have been rescaled (by 0.3 on the left, and by
the factor $K=\sigma_{\sss NLO}/\sigma_{\sss LO}$ on the right).
}
  \end{center}
\end{figure}
%%%%%%%%%%%%%%%%%%%%%%%%%%%%%%%%%%%%%%%%%%%%%%%%%%%%%%%%%%%%%%%%%%%
We begin by considering $t\bar{t}$ production. We have studied, at the
Tevatron and at the LHC, single-inclusive $\pt$ and rapidity spectra of the 
$t$ and $\bar{t}$ decay products, and the correlations in transverse 
momentum, $\Delta\phi$, and invariant mass of the $b\bar{b}$, $l^+l^-$, 
$bl^-$, $\bar{b}l^+$, $b\bar{\nu}$, and $\bar{b}\nu$ pairs. We have
found that angular correlations have an almost negligible impact. 
We present in fig.~\ref{fig:one} the only two observables for which 
these correlations have a visible effect, albeit barely so for $\pt(l^+l^-)$.
On the other hand, angular correlations are an important ingredient for 
the correct prediction of $\Delta\phi(l^+l^-)$, as shown in the right
pane of fig.~\ref{fig:one}. It is interesting that about 30\% of 
the difference between the LO prediction without angular correlations
(dashed histogram -- \HW) and the NLO prediction with angular correlations
(solid histogram, overlayed with open circles -- MC@NLO) is due to beyond-LO
corrections.

\begin{figure}[htb]
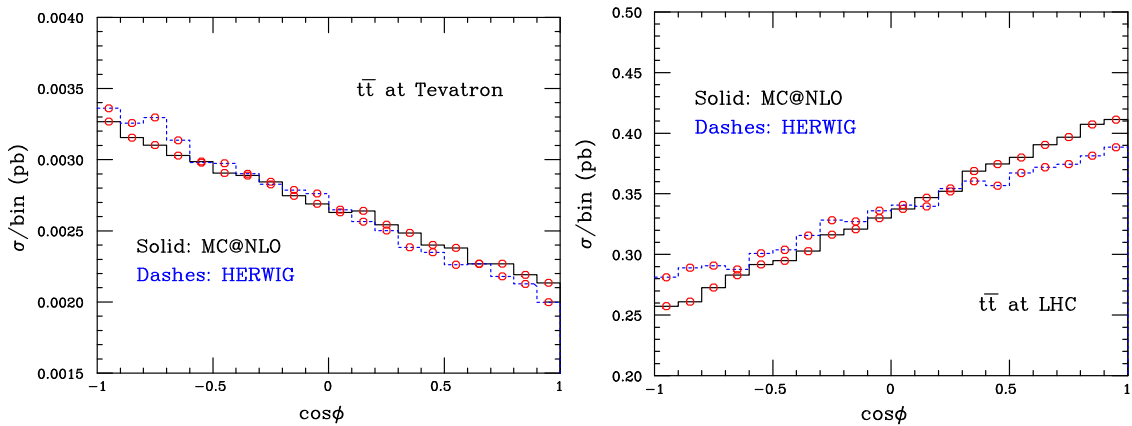

  \begin{center}
    \epsfig{figure=cphi_TEV2.eps,width=0.49\textwidth}
    \epsfig{figure=cphi_LHC2.eps,width=0.49\textwidth}
\caption{\label{fig:two} 
Opening angle distributions, as defined in the text, for $t\bar{t}$ 
production at the Tevatron (left pane) and at the LHC (right pane).
\HW\ results have been rescaled by the $K$ factor. The corresponding
curves obtained by neglecting angular correlations are flat, and
are not shown in the figure.
}
  \end{center}
\end{figure}

It is possible to specifically design observables which would be trivial 
if angular correlations were neglected. Typically, such observables are
angular variables constructed with the decay products of the top quarks, 
and measured in the rest frames of the parent particle. We have considered
the distributions in $\cos\theta_1$, $\cos\theta_2$, and $\cos\phi$,
as defined in ref.~\cite{Bernreuther:2004jv}; in particular,
$\phi$ is the angle between the direction of flight of $l^+$ and the 
direction of flight of $l^-$. The directions of flight are
defined in the $t$ and $\bar{t}$ rest frames respectively
(see ref.~\cite{Bernreuther:2004jv} for more details). 
Results for $\cos\phi$ are presented in
fig.~\ref{fig:two} for the Tevatron (left pane) and the LHC (right pane).
Just as for $\Delta\phi(l^+l^-)$, beyond-LO contributions are not negligible,
and they tend to deplete (at the Tevatron) or to enhance (at the LHC)
the LO predictions for the $\cos\phi$ asymmetry. This behaviour is also 
found in the pure-NLO, parton-level study of ref.~\cite{Bernreuther:2004jv}.
We have verified that, by neglecting angular correlations, the cross
section depends trivially on $\theta_1$, $\theta_2$ and $\phi$.
%%%%%%%%%%%%%%%%%%%%%%%%%%%%%%%%%%%%%%%%%%%%%%%%%%%%%%%%%%%%%%%%%%%
%%%%%%%%%%%%%%%%%%%%%%%%%%%%%%%%%%%%%%%%%%%%%%%%%%%%%%%%%%%%%%%%%%%

Finally, we examine distributions for single-top production
at the Tevatron. Because both production and decay occur through
the left-handed charged current interaction, one expects stronger
production angular correlations than in top quark pair production.
Indeed, angular correlation effects are clearly visible in the single-inclusive
spectra of the top decay products. As in the case of $t\bar{t}$ production,
it is possible to study angular correlations more directly by choosing
specific observables. These observables always involve the definition
of a spin basis that leads to nearly 100\% correlation between the 
direction of the charged lepton from top decay and another 
experimentally-definable, channel-dependent direction~\cite{Mahlon:1996pn}. 
For both $s$- and $t$-channel processes the optimal spin quantization axis 
lies, in the top quark rest frame, along the down-type quark attached
to the vertex connected via a $W$-boson to the top quark producing vertex.
At LO that corresponds for the $s$ channel to the beam-direction, while
for the $t$ channel this is most often the direction of the light quark 
jet against which the top quark recoils. 
%%%%%%%%%%%%%%%%%%%%%%%%%%%%%%%%%%%%%%%%%%%%%%%%%%%%%%%%%%%%%%%%%%%
\begin{figure}[htb]
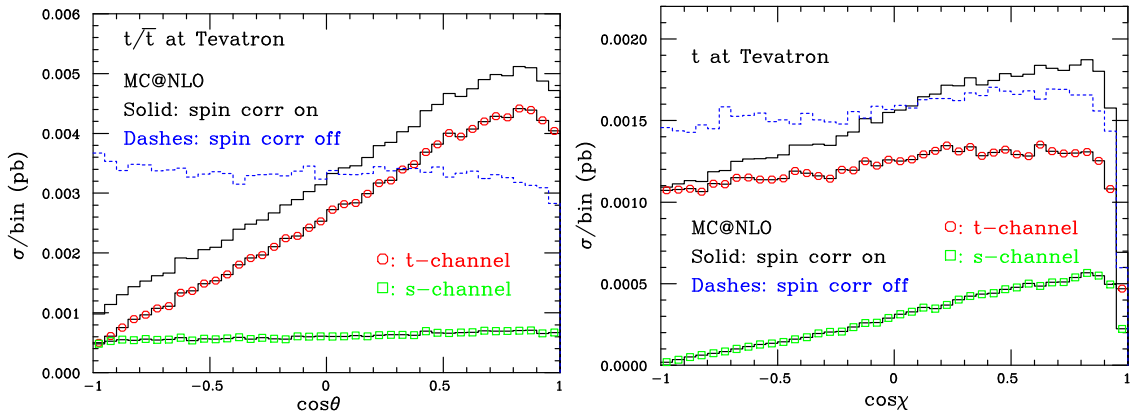

  \begin{center}
    \epsfig{figure=ctheta_st_TEV2.eps,width=0.49\textwidth}
    \epsfig{figure=cxi_st_TEV.eps,width=0.49\textwidth}
\caption{\label{fig:three} 
Angular correlations in single-top processes at the Tevatron:
$\cos\theta$ in single-$t/\bar{t}$ production (left pane),
and $\cos\chi$ in single-$t$ production (right pane). Histograms
without symbols are the sums of $s$- and $t$-channel contributions.
}
  \end{center}
\end{figure}
%%%%%%%%%%%%%%%%%%%%%%%%%%%%%%%%%%%%%%%%%%%%%%%%%%%%%%%%%%%%%%%%%%%

Accordingly, we present in the left pane of fig.~\ref{fig:three} the 
distribution in the cosine of the angle $\theta$, defined as the angle 
between the direction of flight of the lepton emerging from top decay, 
and the axis of the hardest jet which does not contain a stable $b$-flavoured
hadron; the angle is defined in the rest frame of the top quark.
This distribution has been shown in ref.~\cite{Stelzer:1998ni} at
tree level, and in ref.~\cite{Sullivan:2005ar} at NLO using 
MCFM~\cite{Campbell:2004ch}. We have applied similar cuts as those 
in ref.~\cite{Stelzer:1998ni}, namely we required the decay products 
of the top to have
\beqn
&&\pt(b)\ge 20~{\rm GeV}\,,\;\;\;\;\;\;
\abs{\eta(b)}\le 2\,,
\\
&&\pt(l)\ge 10~{\rm GeV}\,,\;\;\;\;\;\;
\abs{\eta(l)}\le 2.5\,,
\\
&&\pt(\nu)\ge 20~{\rm GeV}\,.
\eeqn
We also require the hardest light jet to have transverse momentum larger than 
20 GeV, and $\abs{\eta(j)}\le 2.5$. In this way, we obtain $A=-0.35$, where
\beq
A=\frac{\sigma(-1\le\cos\theta <-0.1)-\sigma(-0.1\le\cos\theta <0.8)}
{\sigma(-1\le\cos\theta <-0.1)+\sigma(-0.1\le\cos\theta <0.8)}\,.
\eeq
As can be seen from fig.~\ref{fig:three}, this result is due to the 
contribution of the $t$-channel, the $s$-channel having a very small asymmetry.
We remark that the asymmetry is also compatible with zero if spin correlations
are switched off. 
It is important to notice that our results follow the same pattern 
(and are actually close numerically) of those of 
refs.~\cite{Stelzer:1998ni,Sullivan:2005ar}. Although we did not carry
out a comprehensive study, this fact implies that not only is the $\cos\theta$
asymmetry fairly robust when including higher order corrections, but it is
also stable when passing from a parton-level description such as that of
refs.~\cite{Stelzer:1998ni,Sullivan:2005ar} to a more realistic hadron-level
description such as that of MC@NLO. 

We conclude by presenting in the right pane of fig.~\ref{fig:three} the
distribution in the cosine of the angle $\chi$, which is defined analogously
to the angle $\theta$, except for the fact that the reference direction is
chosen to be that of the antiproton beam (at variance with the case of
$\cos\theta$, we have limited ourselves here to considering $t$ production,
rather than $t+\bar{t}$ production). As expected~\cite{Mahlon:1996pn},
the dominant contribution to the asymmetry is due in this case to
the $s$-channel. An extremely small non-zero asymmetry may also be
visible in the case in which angular correlations are not included;
we have verified that this is an artifact of the cuts adopted in
the present analysis.

\section{Conclusions\label{sec:concl}}
We have presented a method for the efficient inclusion of 
angular correlations due to production spin correlations in Monte  
Carlo event generators. The method has been demonstrated in detail  
for vector boson and top quark decays, but it is in fact quite  
general, relying only on the fact that the matrix elements do not  
contain sharp features that would lead to unacceptably low  
efficiency. The method is exact, and equivalent to what is
currently implemented in LO-accurate event generators such as
\HW. When the event generator is matched to NLO predictions, as 
is the case for MC@NLO and POWHEG, the resulting correlations are 
correct to LO in soft and collinear regions and to NLO elsewhere.  
The method has been implemented in MC@NLO for $WW$, $t\bar t$ and 
single-top hadroproduction and leptonic decay, and we have presented 
illustrative results for the latter two cases. These results show that  
significant correlations are present in suitably chosen observables.
Version 3.3 of MC@NLO implements off-shell effects only in the case
of $WW$ production. Future versions will include off-shell effects
in top decay; also, vector bosons and top quarks decaying hadronically 
can be simulated using the formalism presented here, bearing in mind 
that NLO corrections to decays are neglected.

\acknowledgments
We would like to thank the CERN TH division for hospitality during the 
completion of this work. We also would like to thank Fabio Maltoni for 
his collaboration during early stages of this work, and Chris White
for useful discussions. The work of E.L. and P.M.  
is supported by the Netherlands Foundation for Fundamental Research of
Matter (FOM) and the National Organization for Scientific Research (NWO);
that of B.W. is supported in part by the UK Particle Physics and Astronomy 
Research Council. 

\appendix
\section{Upper bounds in vector boson production\label{sec:app}}
In this appendix, we present an alternative derivation of eq.~(\ref{Allupp}).
One introduces the quantity
\beq
\Ns^\nu\equiv \Ms^\mu\frac{i}{q^2-\mV^2+i\mV\GammaV}
\left(-g_\mu^{\;\;\nu}+\frac{q_\mu q^\nu}{\mV^2}\right)\;,
\eeq
with which eq.~(\ref{sVBampsq}) becomes
\beq
\sum_{\rm spins}\abs{\As}^2=
F_V^2\,{\rm Tr}\!\left[\left(\vecVcp^2+\axlVcp^2-2\vecVcp\axlVcp\gamma_5\right)
\DiracSlc{k}_1\, \DiracSld{\Ns}\, \DiracSlc{k}_2\, \DiracSld{\Ns}^*\right].
\label{sVBampsqBW}
\eeq
Evaluating this in the rest-frame of the (virtual) vector boson, with the 
$z$-axis along the direction of the lepton 3-momentum, we find
\beq
\sum_{\rm spins}\abs{\As}^2=
2 q^2 F_V^2\,\left[\left(\vecVcp^2+\axlVcp^2\right)
\left(\Ns^1 \Ns^{1*}+\Ns^2 \Ns^{2*}\right)
+4\vecVcp\axlVcp\,{\rm Im} \left(\Ns^1 \Ns^{2*}\right)\right]\;.
\eeq
To establish an upper bound on this quantity, we note that
\beq
2\abs{{\rm Im} \left(\Ns^1 \Ns^{2*}\right)} \leq 
2\abs{\Ns^1}\,\abs{\Ns^2} \leq
\Ns^1 \Ns^{1*}+\Ns^2 \Ns^{2*}
\eeq
and so
\beqn
\sum_{\rm spins}\abs{\As}^2 &\leq&
2 q^2 F_V^2\,\left(\abs{\vecVcp}+\abs{\axlVcp}\right)^2
\left(\Ns^1 \Ns^{1*}+\Ns^2 \Ns^{2*}\right)
\nonumber\\*
&=& 
\frac{2 q^2 F_V^2
\,\left(\abs{\vecVcp}+\abs{\axlVcp}\right)^2}
{(q^2-\mV^2)^2+(\mV\GammaV)^2}
\left(\Ms^1 \Ms^{1*}+\Ms^2 \Ms^{2*}\right)\;.
\eeqn
Note that $\Ms^{1,2}$ in this expression are strictly off-mass-shell
quantities: no on-shell approximations have been made at this stage.

Now consider the production of a stable vector boson of mass $\mV$. 
Denoting the amplitude for this by $\bar \As$, we have (again in the
vector boson rest frame)
\beqn
\sum_{\rm spins}\abs{\bar \As}^2 &=&
\bar \Ms^\mu\bar \Ms^\nu
\left(-g_{\mu\nu}+\frac{q_\mu q_\nu}{\mV^2}\right)_{q^2=\mV^2}\nonumber\\
&=& \bar \Ms^1 \bar \Ms^{1*}+\bar \Ms^2 \bar \Ms^{2*}+\bar \Ms^3 \bar \Ms^{3*}
\eeqn
where $\bar \Ms^\mu$ denotes the on-mass-shell value of $\Ms^\mu$.
Therefore, as long as
\beq\label{MbarM}
\Ms^1 \Ms^{1*}+\Ms^2 \Ms^{2*}\leq
\bar \Ms^1 \bar \Ms^{1*}+\bar \Ms^2 \bar \Ms^{2*}+\bar \Ms^3 \bar \Ms^{3*}
\eeq
we have
\beq
\sum_{\rm spins}\abs{\As}^2 \leq
\frac{2 q^2 F_V^2
\,\left(\abs{\vecVcp}+\abs{\axlVcp}\right)^2}
{(q^2-\mV^2)^2+(\mV\GammaV)^2}
\sum_{\rm spins}\abs{\bar \As}^2
\eeq
and hence
\beq
\frac{d\sigma_{l\bl}}{d\Phi_{2+1^{\!\star}}}\leq
\frac{2 q^2 F_V^2
\,\left(\abs{\vecVcp}+\abs{\axlVcp}\right)^2}
{(q^2-\mV^2)^2+(\mV\GammaV)^2}
\frac{d\sigma_V}{d\Phi_{1+1^{\!\star}}}\;,
\label{Allupp2}
\eeq
which is identical to eq.~(\ref{Allupp}), given the fact that 
$\vecVcp\axlVcp >0$, and that both equations are valid on-shell.

Clearly, one may check whether the bounds given in eqs.~(\ref{Allupp})
and~(\ref{Allupp2}) are not violated in the case of off-shell vector 
bosons. This is indeed the case, provided that the off-shellness is 
not too large or too small (typically, this happens within 
$\pm 30\GammaV$ of the pole mass). A good strategy is that of 
using eq.~(\ref{Allupp}) for $q^2<\mV^2$, and eq.~(\ref{Allupp2})
for $q^2>\mV^2$. However, one should bear in mind that in the case
of off-shell particles the values of Bjorken $x$'s, and hence of the
PDFs, may change, thus potentially affecting the bound.

\end{document}